\documentclass[sigconf,nonacm]{acmart}
\AtBeginDocument{%
  }

\usepackage{placeins}

\usepackage{listings}

\begin{document}

\title{Learning to structure data from user-generated thematic corpora}

% \author{Elishay Avram}
% \email{elishayavram@gmail.com}
% % \email{elishai.avram@live.biu.ac.il}

% \author{Oren Glickman}
% \email{oren.glickman@biu.ac.il}

% \author{Elad Yom-Tov}
% \email{elad.yom-tov@biu.ac.il}

% \affiliation{%
%   \institution{Bar-Ilan University}
%   \city{Ramat Gan}
%   \country{Israel}}

\author{Elishay Avram}
\email{elishayavram@gmail.com}
\affiliation{%
  \institution{Bar-Ilan University}
  \city{Ramat Gan}
  \country{Israel}}

\author{Oren Glickman}
\email{oren.glickman@biu.ac.il}
\affiliation{%
  \institution{Bar-Ilan University}
  \city{Ramat Gan}
  \country{Israel}}

\author{Elad Yom-Tov}
\email{elad.yom-tov@biu.ac.il}
\affiliation{%
  \institution{Bar-Ilan University}
  \city{Ramat Gan}
  \country{Israel}}
\renewcommand{\shortauthors}{Avram et al.}

\begin{abstract}
Thematic corpora, such as social media communities and domain-specific forums, contain large volumes of unstructured text describing data that could be made structured. These include, for example, personal attributes, behaviors, and experiences mentioned in social media data.
Extracting structured representations from such data is challenging, as relevant attributes are often implicitly expressed, are domain-dependent, and are not known in advance.

We propose a fully automated, iterative framework for discovering and extracting domain-specific attribute schemas from thematic corpora without assuming a predefined ontology.
Given a thematic corpus, the framework iteratively induces candidate attributes using Large Language Models (LLMs), sequentially consolidates semantically overlapping attributes, and assigns each attribute a structural type. These enable creating an ontology and populating it with values from the corpus. Furthermore, the framework enables the use of smaller LLMs for value extraction with estimable accuracy loss compared to large LLMs.

We evaluate the proposed framework on 5 health-related Reddit communities. The discovered attribute sets achieved $61\%$ agreement with human-identified attributes, nearly matching the $62\%$ agreement observed between independent annotators. In most cases (3 of 5), the algorithm converges to a stable set of attributes in fewer than 10 iterations, indicating rapid schema stabilization. The framework further assigns structural types to attributes with $82\%$ accuracy, enabling consistent downstream extraction. Once attributes and their structure are established, the framework reaches an $F_{1}$ score of $0.80$ in value extraction compared to human annotations. When evaluating smaller instruction-tuned models against a high-capacity reference LLM, performance improves consistently and statistically significantly with model scale across four LLM families, indicating that increased capacity directly translates into higher extraction quality while enabling informed accuracy--cost trade-offs.

Overall, the results demonstrate that attributes comparable to those identified by humans can be discovered automatically, enabling the creation of high-quality structured datasets economically and at scale. By removing the need for predefined ontologies, iterative model-driven schema induction offers a practical and scalable foundation for mining thematic corpora.
\end{abstract}





\maketitle

\section{Introduction}
Thematic corpora, such as social media communities and domain-specific forums, contain large volumes of unstructured text describing data that could be made structured. These include, for example, personal attributes, behaviors, and experiences mentioned in social media data.
In electronic health records, information such as symptoms, patient history, lifestyle factors, and contextual observations is often recorded in free-text narratives rather than in structured fields. Structuring such information into tabular forms can enable better analysis and utilization of these data (e.g., \cite{sui2024table}).

However, extracting structured data from thematic corpora poses substantial challenges. First, the relevant attributes are rarely known in advance and vary significantly across domains, data sources, and populations. Second, important information is frequently expressed implicitly, using diverse linguistic formulations that resist rule-based or template-driven extraction. Third, most existing information extraction pipelines rely on predefined ontologies or hand-crafted schemas, which require extensive expert effort to design and maintain, and often fail to generalize beyond the specific domain for which they were created. As a result, prior approaches are typically limited to narrow, predefined attribute sets or require substantial human intervention to adapt to new corpora.

Recent advances in Large Language Models (LLMs) offer new opportunities for mining unstructured text with minimal supervision.
However, naively applying LLMs to large corpora introduces new challenges for data mining at scale.
Free-form extraction often yields redundant or unstable attribute sets\cite{minIE}, and repeated inference with large models can incur substantial computational costs.
These limitations motivate the need for schema discovery methods that are both robust and cost-aware, while remaining adaptable across domains.

In this work, we propose an iterative framework for discovering and extracting domain-specific personal attributes from thematic corpora without assuming a predefined ontology.
Given a corpus associated with a specific online community, the framework iteratively induces candidate attributes using instruction-tuned LLMs, consolidates semantically overlapping attributes, and stabilizes the resulting schema across iterations.
Discovered attributes are further assigned structural types, enabling consistent and automated extraction of attribute values across large document collections.

We evaluate the proposed framework on five large Reddit communities spanning diverse health-related domains.
Our experiments analyze schema stability, consolidation strategies, attribute typing accuracy, and the effect of model scale on extraction quality.
Crucially, we demonstrate that effective and stable schema discovery can be achieved using small LLMs, enabling scalable and cost-efficient mining of personal attributes from user-generated text.

\section{Related Work}

\subsection{Information Extraction from Unstructured Text}

Information Extraction (IE) is the process of converting unstructured natural language text into structured representations that can be further analyzed or utilized in downstream tasks~\cite{sarawagi2008information, jiang2012information}. IE is often divided into two main tasks: Attribute discovery and information extraction.

\subsubsection{Automatic Attributes Discovery}

Attribute discovery refers to the task of identifying a meaningful set of attributes that characterize a corpus or dataset, often under conditions of limited prior specification and minimal supervision. Naturally, it is possible to skip this stage if there is a predefined schema that determines the set of attributes to be extracted, thereby constraining the structured representation of information to categories specified in advance. However, in our work, we assume that such a schema does not exist.

Recent work has explored various approaches to attribute discovery, demonstrating the growing interest in methods for identifying informative attributes within unlabeled data.
Zou et al.~\cite{glean} introduced a unified framework that extracts semantic feature representations from textual instances, identifies ambiguous instances, and utilizes LLM feedback to refine representations, generate descriptive group semantics, and align uncertain samples with appropriate data groupings. By incorporating both instance-level and cluster-level feedback, the framework addresses challenges related to limited supervision and semantic representation. Fan et al.~\cite{ligd} proposed an LLM-enhanced framework that integrates LLM annotations with a semantic-correlation graph to support the discovery of previously unseen groups within data. Their method clusters samples, constructs a graph that preserves semantic relationships, and selects high-value instances to be annotated by an LLM. The resulting pseudo-labels are propagated through the graph, and representations are updated through an iterative process until convergence. %Experimental results show improved clustering performance and stronger generalization when identifying unseen data groups under limited supervision.

While these approaches demonstrate the potential of LLMs to support discovery processes, they primarily rely on representation learning and clustering to identify latent data groupings. However, they do not explicitly address the challenge of discovering semantically meaningful attributes directly from unstructured thematic corpora, where attributes must be defined and consolidated without a predefined structure.

% Fan et al. [REF?] propose LLM-Enhanced Generalized Category Discovery via Iterative Graph Diffusion (LIGD), a framework that integrates LLM annotations with a semantic-correlation graph to improve category discovery. The method clusters samples, constructs a graph that preserves semantic relationships, and selects high-value samples—including hard negatives and graph-central instances—to be annotated by the LLM as belonging to either existing or novel categories. The assigned pseudo-labels are then propagated through the graph, and a contrastive learning model updates the representations in an iterative process until convergence. Experimental results show improved clustering performance and stronger generalization when identifying unseen classes under limited supervision.

% Zou et al. \cite{glean} introduced a unified framework that leverages diverse and quality-enhanced LLM feedback for generalized category discovery. Their pipeline first extracts features and mines ambiguous instances, after which LLM feedback is used to refine contrastive representations, generate category descriptions, and align ambiguous samples with appropriate categories. By incorporating both instance-level and cluster-level feedback into training, the framework addresses limitations such as insufficient supervision and the underutilization of semantic meanings in discovered categories. 

\subsubsection{Information Extraction}

Once a set of attributes has been defined or is otherwise provided, information extraction methods can be applied to systematically extract relevant information from unstructured text according to these attributes.

One of the most widely studied subtasks in IE is Named Entity Recognition (NER), which involves identifying and classifying mentions of real-world entities such as persons, organizations, locations, and numerical attributes within raw text~\cite{nadeau2007survey, li2020survey}. NER has been researched extensively over the past two decades, evolving from early rule-based and statistical models (e.g., HMMs, CRFs) to modern neural architectures (e.g., BiLSTM-CRF, transformer-based models) that achieve state-of-the-art performance on benchmark datasets~\cite{lample2016neural, devlin2019bert}. The advent of contextualized language models has significantly improved the ability to learn entity boundaries and types directly from data, reducing dependence on handcrafted features and domain knowledge~\cite{devlin2019bert}.

Traditional IE systems often build upon NER as a first step in a pipeline, followed by relation extraction (RE) to detect semantic relationships between entities, enabling the construction of structured knowledge bases and databases from raw text~\cite{bose2021survey, detroja2023survey}. These pipelines have been applied in various domains, including news, biomedicine, and clinical records, where structured data extraction from narrative text is essential~\cite{newsNER, perera2020ner, clinicalNER}. Advances in deep learning have further transformed NERs, improving the ability of models to learn contextual features and generalize across domains~\cite{yang2022survey}. 

While NER and relation extraction focus on identifying discrete information units and their semantic connections within text, many real-world applications require organizing multiple attributes into unified structured representations that capture complex relationships among them. Beyond extracting individual entities or relations, recent research has therefore explored methods that generate structured representations directly from unstructured text, including formats such as tables, graphs, or other multi-attribute structures. One emerging line of work focuses on \emph{text-to-table} generation, framing the task as the conversion of unstructured narrative text into tabular forms that capture multiple attributes and their interdependencies~\cite{Text-to-Table}.

Thus, Liu et al.~\cite{liu2024structured} developed an LLM-based framework for converting unstructured data into structured representations, enabling scalable processing of diverse textual sources and supporting downstream analytical tasks. Similarly, Rajmohan et al.~\cite{LLMdriventexttotable} proposed an LLM-driven text-to-table system that decomposes the generation process into guided intermediate sub-tasks and iteratively refines the resulting tables through self-feedback. This structured generation strategy demonstrates strong performance on complex public datasets.

While these approaches highlight the effectiveness of LLMs in improving structured extraction and generation, they primarily focus on populating predefined structures. We suggest that schema induction and value extraction
should not be treated as independent stages, but rather as tightly
coupled components of a unified framework. The reason for this is that even after the attribute schema stabilizes, the discovered attributes
often do not align with predefined extraction targets supported by
existing IE systems, as they tend to be domain-specific, corpus-driven,
and semantically nuanced. Applying off-the-shelf extraction methods
would therefore require substantial adaptation, effectively redefining
the extraction process for each newly induced schema. 

Thus, by jointly supporting both processes, the induced schema can directly operationalize the construction of consistent structured datasets without relying on predefined ontologies.

\subsection{Social Media as a Medical Data Source}

Social media platforms such as Reddit have emerged as unexpected goldmines of health-related insights, where people share their firsthand experiences in ways that illuminate the complexities of health. However, the challenge of extracting relevant information from social media is usually addressed in an ad-hoc manner. 

Hochberg and Yom-Tov \cite{yomtov2023large} used regular expressions to extract diet-related information from weight loss fora on Reddit. Similarly, Fennig et al. \cite{fennig2025bridging} extracted information on epilepsy from a Reddit forum, using a predefined schema and an LLM for information extraction.  Rajpurkar et al.~\cite{askdocsLLM} evaluated the effectiveness of LLMs in transforming unstructured posts from the more general \textit{r/AskDocs} subreddit into structured representations. Their methodology combined human annotation with LLM-based extraction, where predefined attributes such as demographic information, medical history, and consultation status were populated using few-shot prompting and templates. However, the proposed framework assumed a predefined schema that determines which attributes should be extracted. 
 
 The challenge, therefore, is to enable information extraction without assuming that the relevant attributes are known in advance.

%As an example, consider Sarker et al. (2020)\cite{MiningCovidSympthomsReddit}, who tapped into this vast resource, analyzing nearly 70,000 posts from the “/r/covidlonghaulers” subreddit. By employing named-entity recognition (NER) and utilizing existing NER models dedicated to medicine, they identified medications and supplements mentioned by users, preprocessing the data to extract relevant entities from the text. 

\begin{figure}[t!]
    \centering
    \includegraphics[
    width=\columnwidth,
    height=0.9\textheight,
    keepaspectratio
]{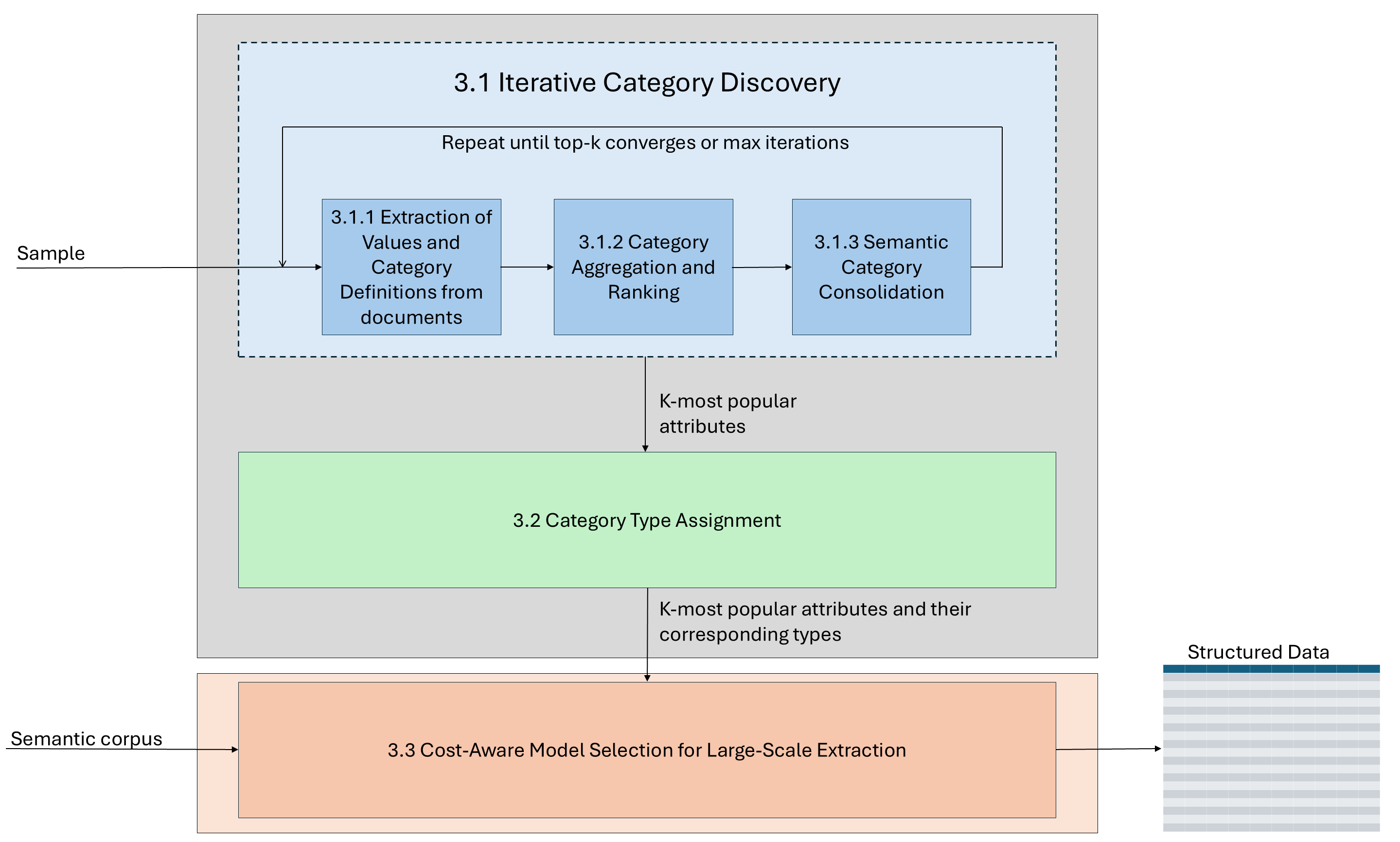}    
% ]{figures/pipeline architecture8.png}

    \Description{ Overview of the proposed framework}
    \caption{
    Overview of the proposed framework for iterative attribute discovery and extraction.
    The process begins with an iterative loop (Section~3.1), in which candidate attributes and values are extracted from documents, aggregated and ranked, and subsequently consolidated using semantic merging.
    This loop repeats until the $k$ most popular attribute set converges or a maximum number of iterations is reached.
    Once a stable set of attributes is obtained, each attribute is assigned a structural type (Section~3.2), followed by cost-aware model selection for large-scale value extraction (Section~3.3).
    }
    \label{fig:pipeline}
\end{figure}
\section{Model}

% \begin{figure*}[t]
%     \centering
%     \includegraphics[width=\textwidth]{figures/Illustarative running example.png}
%     \caption{Illustrative running example of the personal-trait extraction and aggregation process,
% using synthetically generated diet-related posts.
% The figure demonstrates how personal traits are extracted from raw documents, assigned to free-form attributes,
% counted at the document level, and subsequently aggregated and consolidated across the corpus.
% Colors denote different trait attributes.}
%     \label{fig:runningexample}
% \end{figure*}

%\begin{figure*}[t]
%    \centering
%    \includegraphics[width=\textwidth]{figures/pipeline architecture4.png}
%    \Description{ Overview of the proposed framework}
%    \caption{
%    Overview of the proposed framework for iterative attribute discovery and extraction.
%    The process begins with an iterative loop (Section~4.1), in which candidate attributes and values are extracted from documents, aggregated and ranked, and subsequently consolidated using semantic merging.
%    This loop repeats until the $k$ most popular attribute set converges or a maximum number of iterations is reached.
%    Once a stable set of attributes is obtained, each attribute is assigned a structural type (Section~4.2), followed by cost-aware model selection for large-scale value extraction (Section~4.3).
%    }
%    \label{fig:pipeline}
%\end{figure*}

\begin{figure*}[tb]
    \centering
    \includegraphics[width=0.9\textwidth]{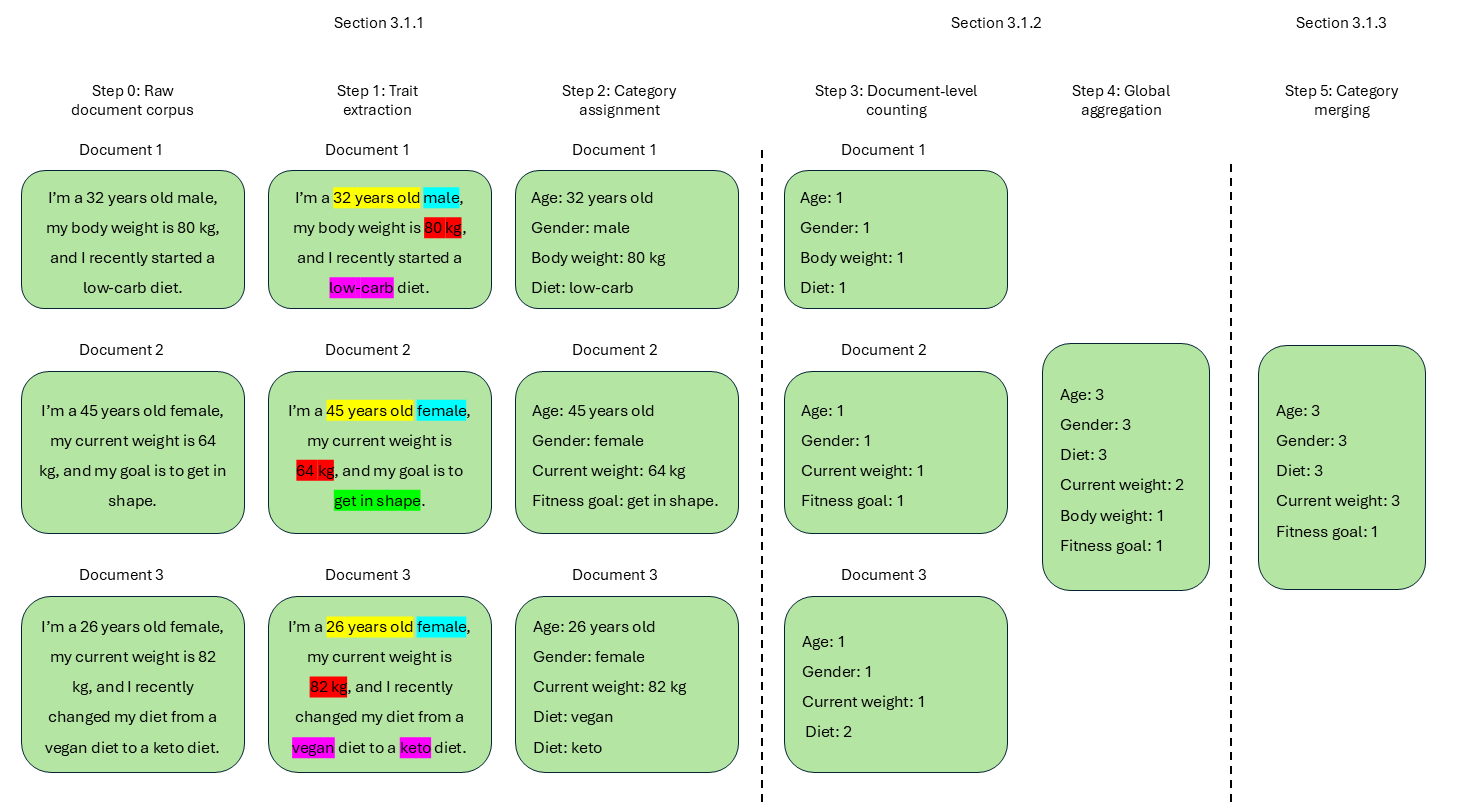}
    \caption{Illustrative example of the extraction and aggregation process described in Section~\ref{sec:iterative_attribute_discovery},
    using synthetically generated diet-related posts.
    The figure demonstrates how personal traits are extracted from raw documents, assigned to free-form attributes,
    counted at the document level, and subsequently aggregated and consolidated across the corpus.
    Colors denote different attributes.}
    \Description{A schematic pipeline showing synthetically generated diet-related posts as input,
    followed by personal-trait extraction, assignment to free-form diet-related attributes,
    document-level counting, and corpus-level aggregation and consolidation.
    Different colors indicate distinct trait attributes throughout the process.}
    \label{fig:runningexample}
\end{figure*}

We propose a fully automated method for transforming free-text content from a thematic corpus into structured data. Given a collection of documents from a thematic corpus, the method automatically identifies shared information attributes that are most prevalent in that corpus and uses them to organize the extracted information into a structured tabular representation.

The resulting table represents the information contained in the corpus, with each row corresponding to a document and each column corresponding to a corpus-specific attribute induced directly from the data. Each table cell contains the value of a given attribute as expressed at the document level, potentially aggregating evidence from multiple mentions within the document. Unlike classical information extraction approaches that focus on identifying and localizing individual attribute mentions in the surface form, our method produces document-level attribute values without requiring explicit span-level annotations.

In the following, we describe the stages of the proposed method. A schematic overview of the method is given in Figure \ref{fig:pipeline}, with reference to the sections below. 

\subsection{Iterative Attribute Discovery}
\label{sec:iterative_attribute_discovery}

The identification of relevant attributes is formulated as an iterative discovery process. Rather than assuming a fixed set of attributes, the method repeatedly refines the attribute set based on information extracted from the data itself. The discovery is based on an analysis of a random sample of $N_{Samples}$ from a thematic corpus with $N_{Corpus}$ documents, where $N_{Corpus}$ >> $N_{Samples}$. 

Each iteration of the process consists of three steps, illustrated through a running example in Figure~\ref{fig:runningexample}:
\begin{enumerate}
    \item Extraction of information and attribute definitions from documents. 
    \item Aggregation and analysis of the extracted attributes.
    \item Semantic consolidation of overlapping or redundant attributes.
\end{enumerate}

The output of each iteration is a set of candidate attributes that serve as the input to the next iteration. The process terminates upon convergence of the attribute set or when a predefined maximum number of iterations ($I_{max}$) is reached.

\subsubsection{Extraction of Information and Attribute Definitions from documents}

At the core of each iteration, the $N_{Samples}$ documents are processed independently by an LLM. For each document, the LLM is prompted to extract information from the document. The type of information that the LLM is prompted to extract is a parameter of the method. For example, the information could be personal, medical, pertaining to an event, etc. The LLM is not constrained with respect to the number of extracted information items and may produce any number of entries per document, including none.

As illustrated in Steps~1--2 of Figure~\ref{fig:runningexample}, for every extracted piece of information, the LLM jointly produces the exact textual value appearing in the document and a free-form attribute name that describes the type of attribute represented by that value. 

The output, therefore, consists of attribute-value pairs grounded directly in the original documents. At this stage, the attribute names are unconstrained, and their value for the same information can vary between documents, reflecting differences in language, granularity, or perspective.

Across iterations, the extraction process maintains a set of supporting evidence drawn from the document corpus and associated with the leading attributes, providing contextual grounding that helps stabilize attribute interpretation over time. The framework is initialized in a zero-shot setting in which the evidence set is empty. Beginning with the second iteration, two textual evidences are retained for each of the $k$ most popular attributes.

\subsubsection{Attribute Aggregation and Ranking}

Following attribute and value extraction, the frequency of each attribute across documents is accumulated. This aggregation step is necessary to move from isolated, document-level extractions to a corpus-level representation, in which attribute prevalence reflects how repeatedly an attribute is expressed throughout the data. Accumulated frequencies provide a quantitative signal that supports ranking attributes and distinguishing dominant attributes from those that appear only sporadically.

Attributes can then be aggregated and ranked based on their corpus-level frequency, allowing the method to retain only the most prevalent attributes for subsequent consolidation.

The aggregation process is illustrated in the running example (Figure~\ref{fig:runningexample}) through the relevant stages, including document-level counts (Step~3) and corpus-level aggregation (Step~4).

\subsubsection{Semantic Attribute Consolidation}

Because attribute names are generated by the LLM in an unconstrained manner, multiple attribute names may refer to the same underlying semantic concept. To address this, each iteration includes a semantic attribute consolidation step, illustrated in Step~5 of Figure~\ref{fig:runningexample}.

Only attributes that appear at least $c_{\text{frequency}}$ times are considered for consolidation, where $c_{\text{frequency}}$  is a pre-set configuration parameter defining the minimum frequency threshold of the method. This restriction limits the consolidation process to attributes with sufficient empirical support, reducing unnecessary computation of attributes with low prevalence.

To select an effective semantic consolidation mechanism, we evaluated four alternative strategies that differ in how attribute similarity is assessed and how merging is applied. Each strategy operates on the set of attributes identified within an iteration and relies on LLM-based semantic judgments to determine attribute equivalence. The four strategies are:

\noindent \textbf{Pairwise comparison}: Every unordered pair of attributes is independently evaluated for semantic equivalence. For each pair, the LLM  considers the attribute names and indicates whether they refer to the same semantic concept. Semantically equivalent attributes are merged. The LLM-call complexity of this strategy is $O(n^2)$, where $n$ is the number of attributes.   

\noindent \textbf{Global grouping}: The full list of attributes is provided to the LLM in a single prompt. The model is instructed to group semantically equivalent attributes into clusters, with each cluster representing a consolidated attribute. This strategy relies on the model’s ability to jointly reason over the entire attribute set in a single consolidation step. The LLM-call complexity of this strategy is $O(1)$.

\noindent \textbf{Incremental consolidation}: Attributes are processed incrementally in descending order of empirical frequency. At iteration $i$, the $i$-th attribute in this ordering is compared to all previously retained attributes $\{1,\ldots,i-1\}$. The LLM considers the attribute names and indicates whether they refer to the same semantic concept. The $i$-th attribute is either merged into an existing attribute or retained as a new distinct attribute. The LLM-call complexity of this strategy is $O(n)$, where $n$ is the number of attributes.   

\noindent \textbf{Bottom-up merging}: An agglomerative merging procedure that starts with the complete set of attributes and iteratively merges attribute sets that the LLM indicates are semantically similar. 
    This approach resembles hierarchical agglomerative clustering \cite{hart2001pattern}, with semantic similarity assessed using LLM-based judgments rather than fixed distance metrics. The LLM-call complexity of this strategy is $O(n)$, where $n$ is the number of attributes.

In our experiments, we evaluated the 4 strategies on one corpus, selected the most accurate, and applied it to all other corpora. This was done to reduce the likelihood of overfitting the strategy to the specific corpora analyzed. 

\paragraph{Stabilization and Final Attribute Set}

The attribute set obtained at the end of each iteration defines the initialization for the subsequent iteration. 
%Attributes are ranked by their empirical frequency, and only the ${k}$ most popular attributes are retained, where ${k}$ is a predefined parameter.
% Starting from the second iteration, the retained ${k}$ most popular attributes are augmented with representative evidence examples extracted from the corpus. This evidence consists of concrete value–context pairs previously identified for each attribute and is incorporated into the prompt as illustrative examples (that is, the LLM prompt uses few-shot conditioning).
The $k$ most popular attributes, where $k$ is predefined, are incorporated into an evidence-augmented prompt that provides representative conditioning signals for the extraction process.
For example, if Figure~\ref{fig:runningexample} illustrates the first iteration, a random evidence instance is selected for inclusion in the subsequent prompt; in this case, one could observe an example such as "Age: 45 years old".

%The evidence used for prompting is derived exclusively from the ${k}$ most popular attributes obtained in the preceding iteration. This design allows the language model to refine attribute extraction and assignment based on empirically dominant patterns observed in the sample data, while preventing information leakage from lower-support or unstable attributes.

Using this evidence-augmented prompt, the extraction, aggregation, and consolidation steps are repeated, allowing attribute definitions and assignments to evolve across iterations in a controlled and data-driven manner.

The iterative process terminates when the set of ${k}$ most popular attributes remains unchanged between consecutive iterations, indicating convergence, or when a predefined maximum number of iterations is reached. Upon termination, the resulting ${k}$ most popular attribute set is treated as the final representation of the dominant personal information dimensions relevant to the thematic corpus.

\subsection{Attribute Type Assignment}

After the discovery of a stable set of thematic corpus-specific attributes, the method determines the structural type of information that each attribute is intended to capture. This step is required in order to support consistent and automated extraction of values across documents, as different attributes correspond to different output structures.

In our framework, each attribute is assigned to one of three predefined attribute types:

\begin{itemize}
    \item Exact value: Attributes whose values are expressed as a single atomic value (e.g., age, height, gender).
    \item List: Attributes that may contain multiple discrete values within a single document (e.g., medications, blood pressures, weights).
    \item Free text: Attributes whose values are expressed as open-ended text without a fixed structure (e.g., emotional state, personal goals).
\end{itemize}

The assignment of attribute types is performed automatically using an LLM. For each attribute discovered, the model is prompted to classify the expected structure of the values associated with that attribute, given the context of the thematic corpus in which it appears. The output is a single attribute type label for each attribute.

\subsection{Cost-Aware Model Selection for Large-Scale Extraction}
\label{sec:cost_aware_model_selection}

After completion of the previous steps, it is possible to utilize a large LLM on the full corpus and structure the information therein into the discovered attributes. It is noted that a large LLM is required due to the complexity of the prompts. However, applying a large LLM to extract information from all documents of a large thematic corpus can be costly. To mitigate this, our framework supports attribute-level model selection, enabling informed trade-offs between extraction accuracy and computational cost.

Once attributes and their types are determined, extraction performance can be evaluated per attribute by comparing the outputs of smaller LLMs against those produced by a large (high-capacity) reference LLM. This comparison provides an estimate, for each attribute, of how closely a smaller LLM approximates the behavior of a larger LLM in terms of extraction quality.

Based on these per-attribute evaluations, users may define accuracy thresholds with respect to a larger reference LLM that determine the smallest LLM that can be used for the extraction of each attribute. For attributes where a smaller LLM achieves performance above a predefined threshold, extraction can be delegated to that LLM, yielding faster inference and lower cost. Attributes that fail to meet the threshold can instead be processed using a larger LLM when higher precision is required.

% Importantly, this mechanism operates at the attribute level rather than the document level, allowing consistent and predictable extraction behavior across a large corpus. The framework itself remains model-agnostic and does not mandate a specific reference model or threshold, enabling flexible adaptation to different application constraints and budgets.

\section{Experimental Settings}

We evaluate our iterative attribute discovery framework on thematic corpora derived from Reddit, a popular social network. In our experimental setup, a \emph{thematic corpus} is operationalized as a single Reddit community (a \emph{subreddit}), representing a coherent domain in which users discuss related topics and exhibit similar patterns of information disclosure. Within each thematic corpus, \emph{documents} correspond to individual Reddit posts or comments authored by users.

Experiments are conducted on five different subreddits. In this work, we focused on medical and health-related subreddits, as follows:

\noindent \textbf{r/TryingToConceive (TTC)}: Subreddit focusing on fertility, conception attempts, and lifestyle factors related to pregnancy.

\noindent \textbf{r/Diabetes}: Subreddit centered on diabetes management, including glucose monitoring, medication usage, diet, and long-term health indicators.

\noindent \textbf{r/Hypertension}: Subreddit covering discussions on blood pressure, cardiovascular health, medication adherence, and lifestyle interventions.

\noindent \textbf{r/Steroids}: A subreddit characterized by detailed self-reports on performance-enhancing drug usage, dosage protocols, physiological measurements, and side effects thereof.

\noindent \textbf{r/AskDocs}: A general medical advice forum where users describe symptoms, medical history, and demographic information when seeking help from medical professionals.

Each subreddit is treated as an independent thematic corpus, allowing attribute discovery to be performed separately for each corpus while preserving a consistent experimental configuration.

For each subreddit-based thematic corpus, we randomly sample $N_{Samples} = 1000$ documents. These documents (posts or comments) serve as the input to all stages of the iterative discovery process. To restrict the attribute set to popular attributes, we retain only the $k = 10$ most popular attributes.

The iterative discovery process is executed with the maximum number of iterations set to $I_{max}=10$.
The minimum frequency threshold for semantic consolidation is set to $c_{\text{frequency}} = 3$.

\subsection{Evaluation Protocol}
\label{sec:evaluation_protocol}

Evaluation of the values extracted for each attribute was conducted at the attribute level, with the comparison procedure adapted to the structural type assigned to each attribute. This design allows the evaluation metric to reflect the expected form of the extracted information. Specifically, for exact value attributes, predictions are considered correct if the model output exactly matches the reference annotation. In list attributes, precision and recall are computed over the sets of extracted values and combined into an F$_1$ score. Finally, for free text, semantic similarity is assessed using an LLM-based comparison. Specifically, the model-generated output and the reference annotation are provided to a large language model, which assigns an integer similarity score between 0 (not similar at all) and 10 (semantically identical). The score is normalized by dividing by 10, yielding a value in the range $[0,1]$.

%\begin{itemize}
%    \item \textbf{Exact value}: Predictions are considered correct if the model output exactly matches the reference annotation.
%    \item \textbf{List}: Precision and recall are computed over the sets of %extracted values and combined into an F$_1$ score.
%    \item \textbf{Free text}: Semantic similarity is assessed using an LLM-based comparison.
%    Specifically, the model-generated output and the reference annotation are provided to a large language model, which assigns an integer similarity score between 0 (not similar at all) and 10 (semantically identical).
%    The score is normalized by dividing by 10, yielding a value in the range $[0,1]$.
%\end{itemize}

\subsection{Large Language Models}

All experiments were performed using \texttt{ChatGPT-4o} as the primary large language model~\cite{openai_gpt4o_docs}.
The model was used with a fixed configuration across all experimental settings, including a temperature of 0 and a maximum generation length of 2048 tokens. We note that the exact parameter count of \texttt{ChatGPT-4o} is not publicly disclosed, but prior estimates report a wide range of possible model sizes, varying from approximately 200B to 1.8T parameters~\cite{explodingtopics_gpt_parameters2}.

We evaluate the use of the proposed framework for information extraction with smaller LLMs using instruction-tuned LLMs from three model families: Llama (1B, 3B and 8B parameters), Qwen (0.5B, 1.5B, 3B, 7B, and 14B parameters), and Mistral (7B parameters), in order to examine the computational trade-offs of iterative attribute discovery. All models are used in an instruction-following setting with identical prompts and without task-specific fine-tuning.

\subsection{Comparison to Human Performance}

\subsubsection{Attribute Discovery}
Two people independently read 20 documents from each subreddit and were asked to propose the 10 most common categories therein using instructions similar to those given in the algorithm prompt. These categories were compared to those discovered by the proposed algorithm. 

\subsubsection{Attribute Type}

To establish a ground truth for attribute type assignment, two human annotators manually annotated the structural type (exact value, list, or free text) of each of the 10 discovered attributes.

\subsubsection{Document-Level Value Annotation}

Following attribute and attribute-type annotation, two human annotators manually annotated 50 randomly sampled documents for each of the attribute values. During annotation, the annotator was provided with the discovered attributes and their corresponding attribute types.
For each sampled document, the annotator manually filled in reference values for all attributes associated with the subreddit, resulting in a complete attribute-level annotation for every document.
These annotations serve as ground truth for evaluating value extraction performance.

\section{Results}

\subsection{Inter-Annotator Agreement}

\subsubsection{Attribute Sets Agreement}

The two human annotators who suggested categories for each subreddit proposed, on average, 62\% similar attributes. The categories they proposed were matched to, on average, 61\% of the categories identified by the proposed algorithm. 

\subsubsection{Document-level Values Annotations Agreement}

%Table~\ref{tab:iaa_hypertension} presents attribute-level agreement levels between two human annotators for the \textit{r/Hypertension} corpus, which is used as a representative example to illustrate agreement patterns at the attribute level. 
Across domains, the inter-annotator agreement scores reported here were computed as detailed in Section~\ref{sec:evaluation_protocol}. The average agreement was 0.71, with agreement levels of 1.00 for exact value attributes, 0.78 for lists, and 0.45 for free text.

%\begin{table}[t]
%\centering
%\caption{Inter-annotator agreement for attribute type assignment in the %\textit{r/Hypertension} corpus.}
%\label{tab:iaa_hypertension}
%\begin{tabular}{lc}
%\hline
%\textbf{Attribute} & \textbf{Agreement} \\
%\hline
%Blood pressure & 0.80 \\
%Health condition & 0.60 \\
%Symptom & 0.70 \\
%Medication & 1.00 \\
%Age & 1.00 \\
%Diet & 0.20 \\
%Gender & 1.00 \\
%Exercise routine & 0.20 \\
%Height & 1.00 \\
%Current occupation & 0.80 \\
%\hline
%\end{tabular}
%\end{table}

%\begin{table}[t]
%\centering
%\caption{Average inter-annotator agreement by attribute type across all thematic corpora.}
%\label{tab:iaa_by_type}
%\begin{tabular}{lc}
%\hline
%\textbf{Attribute Type} & \textbf{Average Agreement} \\
%\hline
%Exact value & 1.00 \\
%List & 0.78 \\
%Free text & 0.45 \\
%\hline
%\end{tabular}
%\end{table}

% \begin{figure*}[t]
%     \centering
%     \includegraphics[width=\textwidth]{figures/final top 10 attributes in 5 subreddits.jpg}
%     \caption{Final top-$k$ attributes discovered for each thematic corpus after convergence ($k=10$).}
%     \label{fig:attributes}
% \end{figure*}

\begin{table*}[tb]
\centering
\caption{The 10 most popular discovered attributes for each thematic corpus after convergence.}
\label{tab:top10_attributes}
\begin{footnotesize}
\begin{tabular}{p{0.18\textwidth} p{0.18\textwidth} p{0.18\textwidth} p{0.18\textwidth} p{0.18\textwidth}}
\hline
\textbf{TTC} & \textbf{Diabetes} & \textbf{Hypertension} & \textbf{Steroids} & \textbf{AskDocs} \\
\hline
Supplements & A1c level & Blood pressure & Medication & Symptom \\
Tracking methods & Medical condition & Health condition & Health condition & Health condition \\
Symptom & Medication & Symptom & Future plan & Age \\
Emotional state & Dietary habit & Medication & Concern & Medication \\
Health condition & Emotional state & Age & Blood test result & Gender \\
Birth control history & Weight loss & Diet & Current activity & Medical test \\
Age & Age & Gender & Age & Weight \\
Community involvement & Exercise habit & Exercise routine & Fitness goal & Height \\
Partner’s age & Gender & Height & Weight & Substance use \\
Partner’s health condition & Cholesterol level & Current occupation & Health supplement & Coping mechanism \\
\hline
\end{tabular}
\end{footnotesize}
\end{table*}

\subsection{Most Popular Attributes Across Thematic Corpora}

Table~\ref{tab:top10_attributes} presents the $k$ most popular attributes discovered for each thematic corpus after convergence.
Across all domains, the resulting attribute sets reflect the thematic focus of the corresponding subreddit. Medical-condition–oriented communities such as \textit{r/Diabetes} and \textit{r/Hypertension} are characterized by clinical attributes, including measurements, diagnoses, medications, and lifestyle-related factors.
In contrast, \textit{r/Steroids} emphasizes goal-driven and activity-related attributes, such as fitness goals, future plans, and supplementation, alongside medical monitoring.
The \textit{r/TryingToConceive} corpus highlights a broader set of relational values, emotional states, and reproductive attributes.
Finally, the more heterogeneous \textit{r/AskDocs} domain yields a diverse attribute set combining symptoms, demographic attributes, medical tests, and behavioral factors.
Overall, the discovered $k$ most popular attributes capture meaningful information that aligns with the goal of each subreddit, despite being extracted without predefined schemas. The final attribute sets
achieved a 61\% agreement with human-identified attributes, indicating
a substantial overlap between the automatically discovered attributes
and those perceived as salient by human annotators.

\subsection{Semantic Consolidation Strategy}

We evaluated four semantic consolidation strategies on the \textit{r/LoseIt} subreddit in order to select a single strategy for use across all thematic domains.
The \textit{r/LoseIt} community is a large health-focused Reddit forum centered on weight loss and lifestyle change.

To avoid biasing the consolidation strategy toward the characteristics of the corpora used in the main experiments, this evaluation was conducted on a held-out thematic corpus.
The \textit{r/LoseIt} subreddit is not included in the downstream experimental evaluation and is used exclusively for consolidation strategy selection.

Among the evaluated methods, the \emph{Incremental consolidation} strategy achieved the best overall performance, with an F$_1$ score of 0.32 (precision 0.25, recall 0.45).
In comparison, the \emph{Bottom-up merging} strategy attained an F$_1$ score of 0.13 (precision 0.12, recall 0.15), \emph{Global grouping} achieved an F$_1$ score of 0.08 (precision 0.25, recall 0.05), and the \emph{Pairwise comparison} approach yielded the lowest performance with an F$_1$ score of 0.08 (precision 0.06, recall 0.10).

%In relative terms, the \emph{Incremental consolidation} strategy outperformed \emph{Bottom-up merging} by a factor of approximately 2.4$\times$, \emph{Global grouping} by nearly 4$\times$, and \emph{Pairwise comparison} by over 4$\times$ in terms of F$_1$ score.

Based on this consistent relative advantage, the \emph{Incremental consolidation} strategy is selected and applied to all other subreddits.

\subsection{Convergence Behavior Across Domains}

We analyze the convergence behavior of the iterative attribute discovery process by measuring the number of iterations required for the $k$ most popular attribute set to stabilize in each thematic corpus.
Convergence is defined as the point at which the discovered set of $k$ most popular attributes remains unchanged, or when a maximum of 10 iterations is reached.

Across the evaluated subreddits, the number of iterations required for convergence varies by domain.
The process converges after 4 iterations for \textit{Steroids}, 5 iterations for \textit{Hypertension}, and 8 iterations for \textit{AskDocs}.
For \textit{TTC} and \textit{Diabetes}, the process reaches the predefined maximum of 10 iterations without earlier stabilization.

In addition, the held-out \textit{r/LoseIt} corpus used for selecting the semantic consolidation strategy converges after 3 iterations.
This observation further illustrates that the iterative process typically stabilizes within a small number of iterations, with domain-specific variation.

\subsection{Attribute Type Assignment Accuracy}

We evaluate the accuracy of the automatic attribute type assignment by comparing model predictions to manual annotations across all discovered attributes.
Overall, the automatic assignment achieves an accuracy of 82\% over the full set of 50 evaluated attributes.
This result indicates that the model is able to infer attribute structure in a domain-aware manner, supporting reliable downstream value extraction without reliance on manually defined rules or domain-specific heuristics.

\subsection{Model-Scale Results}

\begin{figure*}[t]
\centering
\includegraphics[width=0.75\linewidth]{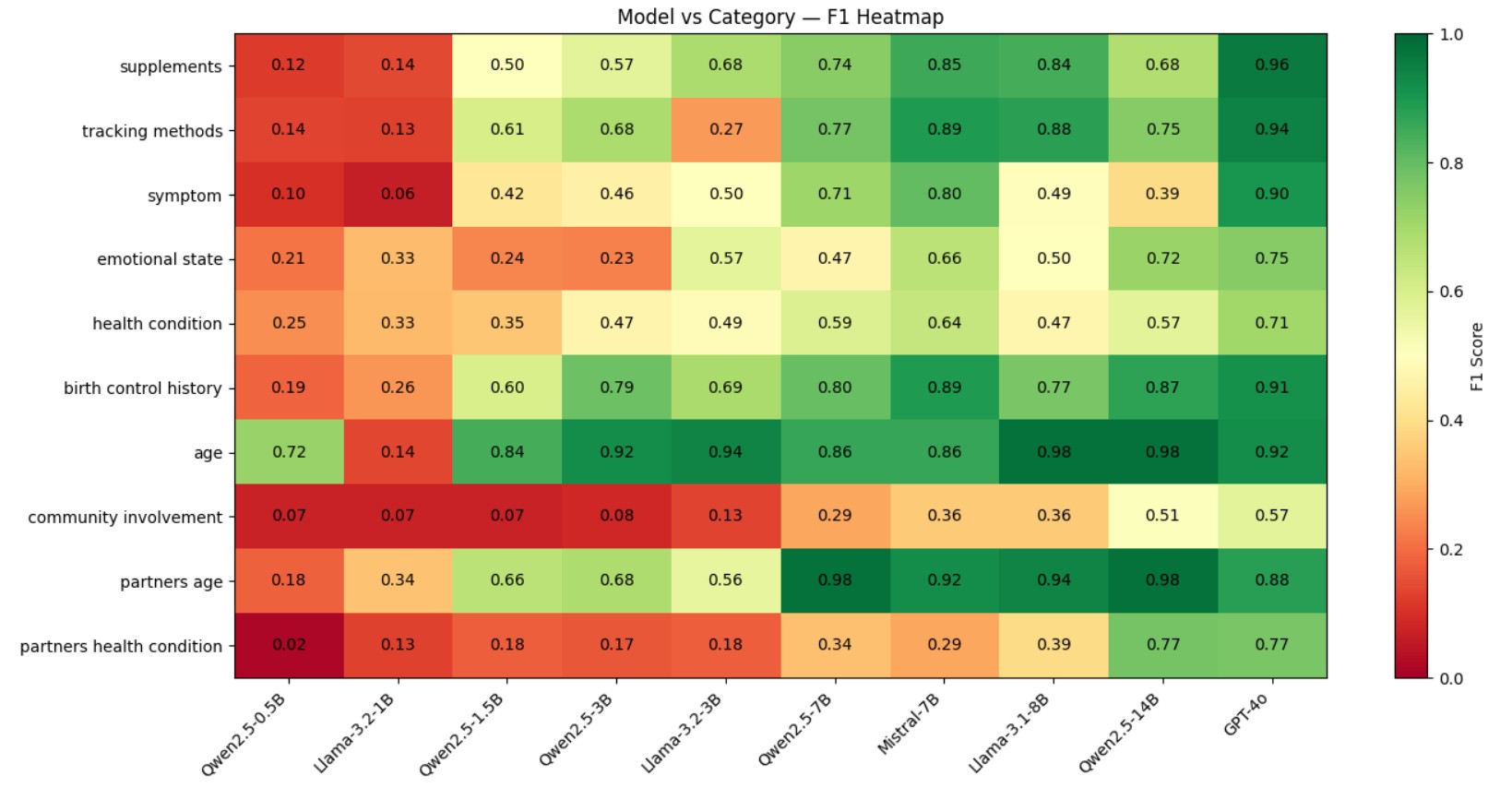}
\caption{Heatmap of attribute-level F$_1$ scores for the \textit{r/TTC} subreddit for  different LLMs.
Rows correspond to the ten most popular discovered attributes, and columns correspond to LLMs with an increasing number of parameters.}
\Description{A heatmap visualizing attribute-level F$_1$ scores for the r/TTC corpus.
Each row represents one of the ten discovered personal-trait attributes, and each column represents a language model ordered by increasing parameter scale.
Cell colors indicate relative performance levels.}
\label{fig:ttc_heatmap}
\end{figure*}

Figure~\ref{fig:ttc_heatmap} presents a heatmap of attribute-level F$_1$ scores for the TTC corpus across the evaluated LLMs.
Rows correspond to the ten final attributes discovered for this domain, while columns correspond to LLMs ordered by increasing number of parameters.
The figure highlights the variability in accuracy across both attributes and model sizes within a single thematic corpus.

\begin{figure}[t]
    \centering
    \includegraphics[width=0.75\linewidth]{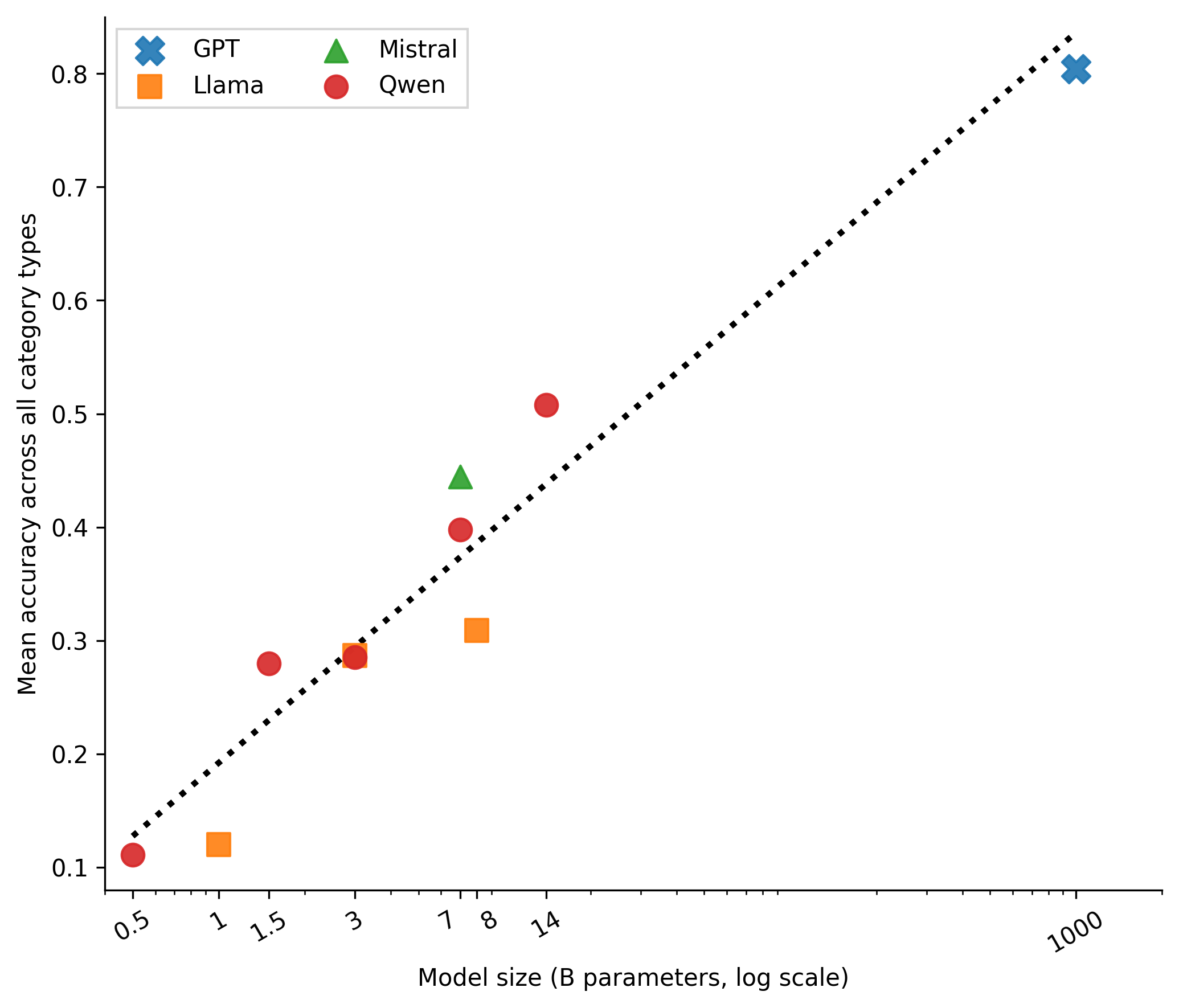}
    \caption{Mean accuracy across all category types as a function of model size (log scale).
    Each marker represents a single model, and the dotted line denotes a log-linear regression fitted across all models.
    The computation of accuracy is described in Section~4.1.}
    \Description{Scatter plot of mean accuracy versus model size on a logarithmic x-axis.
    Markers denote different model families (Qwen, Llama, Mistral, GPT), and a dotted line indicates the fitted log-linear trend.}
    \label{fig:model_size_vs_accuracy_log}
\end{figure}

Figure \ref{fig:model_size_vs_accuracy_log} shows the average accuracy as a function of model family and size, showing the increase in accuracy as a function of model size, while also demonstrating the variability among model families. 

%While Figure~\ref{fig:ttc_heatmap} illustrates detailed attribute-level behavior for one representative domain, aggregated results across all thematic corpora are reported in Tables~\ref{tab:avg_by_model_size} and~\ref{tab:avg_by_model_family}.
%Table~\ref{tab:avg_by_model_size} summarizes average F$_1$ scores as a function of model size, computed by averaging over all attributes and all subreddits.
%Complementarily, Table~\ref{tab:avg_by_model_family} reports average F$_1$ scores aggregated by model family, grouping models according to their architectural lineage.

%Together, Figure~\ref{fig:ttc_heatmap} and Tables~\ref{tab:avg_by_model_size} and~\ref{tab:avg_by_model_family} provide complementary views of model performance across scales and domains.

% \begin{figure}[t]
%     \centering
%     \includegraphics[width=\columnwidth]{figures/model_size_vs_mean_accuracy_logx_regression_03_02.png}
%     \caption{Accuracy vs.\ model size. Markers denote model families, and the black curve shows the mean accuracy per model size (computed over non-GPT families). A broken x-axis is used to display the proprietary GPT result (1000B) without compressing the open-source scales; the dashed arrow indicates the continuation from the largest open-source size to GPT.}
%     \label{fig:model_size_vs_accuracy}
% \end{figure}

% \begin{table}[t]
% \centering
% \caption{Average F$_1$ score by model size, aggregated across all subreddits and attributes.}
% \label{tab:avg_by_model_size}
% \begin{tabular}{lc}
% \hline
% \textbf{Model Size (B)} & \textbf{Average F$_1$} \\
% \hline
% 0.5  & 0.111 \\
% 1.0  & 0.120 \\
% 1.5  & 0.280 \\
% 3.0  & 0.286 \\
% 7.0  & 0.421 \\
% 8.0  & 0.309 \\
% 14.0 & 0.508 \\
% GPT-4o & 0.804 \\
% \hline
% \end{tabular}
% \end{table}

%\begin{table}[t]
%\centering
%\caption{Average F$_1$ score by model family, aggregated across all subreddits and attributes.}
%\label{tab:avg_by_model_family}
%\begin{tabular}{lc}
%\hline
%\textbf{Model Family} & \textbf{Average F$_1$} \\
%\hline
%GPT     & 0.804 \\
%Mistral & 0.444 \\
%Qwen    & 0.316 \\
%Llama   & 0.239 \\
%\hline
%\end{tabular}
%\end{table}

%\subsection{Regression Analysis of Aggregate Performance}

%\begin{figure*}[t]
%    \centering
%    \includegraphics[width=\textwidth]{figures/regression result.jpg}
%    \caption{regression results}
%    \Description{regression results}
%    \label{fig:regression}
%\end{figure*}

To examine how model capacity and structural factors jointly relate to extraction accuracy, we modeled the accuracy using a rank regression model where each observation corresponds to a single attribute-level evaluation, with accuracy serving as the dependent variable. The independent variables are model scale (log-transformed), model family, subreddit, and attribute type. We note that, as the exact parameter count of \texttt{ChatGPT-4o} is not publicly disclosed, we repeated the regression analysis under three plausible estimates of its scale: 200B, 1T, and 1.8T parameters. Across all three assumptions, the estimated coefficients, their statistical significance, and the overall explanatory power of the model remain similar. We therefore report results for the regression based on the intermediate estimate of 1T parameters, which serves as a representative approximation.

%The regression jointly accounts for model scale (expressed in logarithmic form), model family, subreddit domain, and attribute type.
%Model size is log-transformed to reflect the wide and uneven spacing between evaluated models, which span multiple orders of magnitude.

The resulting model achieves $R^2=0.67$. Several predictors exhibit statistically significant associations with accuracy ($p < 0.05$).
Model scale is positively correlated ($\beta = 0.107$, $p < 0.001$), indicating consistent accuracy gains as model capacity increases. Both list-based attributes and number-based attributes are associated with higher accuracy relative to the reference attribute (free-text type) type ($\beta = 0.057$, $p = 0.003$ for the former and $\beta = 0.306$, $p < 0.001$ for the latter). Both Mistral ($\beta = 0.171$, $p = 0.002$) and Qwen ($\beta = 0.136$, $p = 0.015$) are associated with statistically significantly higher accuracy compared to the reference model family (GPT), whereas Llama is not statistically significantly different from GPT.

Finally, among the evaluated domains, the TTC subreddit exhibits statistically significantly better accuracy ($\beta = 0.266$, $p < 0.001$), while coefficients for the remaining subreddits are not statistically significant from the baseline subreddit.

%Overall, these results provide an aggregated quantitative view of how model scale, architectural lineage, attribute structure, and domain characteristics jointly relate to extraction performance across all evaluated settings.

\section{Discussion}

The growing availability of thematic corpora, particularly in online communities, presents a significant opportunity for transforming unstructured narratives into structured data suitable for large-scale analysis. However, this transformation remains inherently challenging, as relevant attributes are often implicit, domain-dependent, and unknown in advance.

To address this challenge, we proposed a fully automated iterative framework for discovering and extracting domain-specific attribute schemas without relying on a predefined ontology. By enabling schema induction directly from the data and stabilizing it through semantic consolidation, the framework supports scalable and cost-aware information extraction across diverse domains.

The experimental findings show the effectiveness of the proposed approach. The discovered attribute sets achieved 61\% agreement with human-identified attributes, closely matching the 62\% agreement observed between independent annotators. Building on the successful identification of domain-relevant attributes, the framework automatically assigns a structural type to each attribute, achieving 82\% accuracy and enabling consistent downstream extraction. Once attributes and their structure are established, the framework achieves an F1 score of 80\% in value extraction relative to human annotations, demonstrating that high-quality structured data can be generated through a fully automated pipeline, where both schema discovery and value extraction operate without human intervention and without relying on predefined schemas.

To better understand the tradeoffs between model capacity and extraction quality, our method evaluates a range of smaller instruction-tuned models and compares their performance to that of GPT-4o, which served as the high-capacity reference model. This comparison enables practitioners to define an accuracy threshold in advance and determine whether a smaller model can be reliably deployed for extraction. 

Our analysis (using a regression model) shows that the accuracy of extraction grows with model size, while highlighting the differences in models: both Mistral and Qwen exhibit better performance than a (theoretically similarly sized) GPT model, whereas Llama does not demonstrate a statistically significant difference. These findings suggest that, under comparable scaling conditions, certain model families may be inherently better suited for structured extraction tasks than others. 

While the proposed framework demonstrates strong performance, several directions remain open for further investigation. First, the current evaluation was conducted on a specific collection of thematic corpora. Extending the analysis to additional domains would provide a broader understanding of the framework’s generalizability and help characterize its behavior under varying linguistic structures and information patterns.

Thus, we plan to apply our framework to the entire Reddit corpus, with the hope of structuring it in a way that enables research. 
%Although the pipeline operates in a fully automated manner, several design parameters were fixed in the present study. Treating these components as tunable hyperparameters presents a promising avenue for future research. In particular, systematically varying the number of sampled documents, the number of retained attributes, and the maximum number of discovery iterations could yield deeper insight into the sensitivity, stability, and convergence dynamics of the iterative process. Such analyses may help establish practical guidelines for configuring the framework across datasets of different scales and characteristics.

%Another important direction involves broadening the range of reference models used during evaluation. GPT-4o served as the high-capacity reference model; however, comparing the framework against additional frontier language models could provide a more comprehensive understanding of how architectural differences influence schema discovery and value extraction.

\section{Ethical Considerations}

This work analyzes user-generated content from health-related Reddit communities, where users may disclose sensitive personal and medical information. Although our framework operates at the corpus level and is intended to discover recurring information attributes and construct structured representations, automatically extracting and organizing such information may increase privacy risks compared with leaving the information in its original unstructured form. In particular, combining multiple extracted attributes could facilitate profiling or reveal sensitive characteristics that users did not intend to expose in a structured dataset.

The proposed framework should therefore be applied with appropriate safeguards when processing user-generated data, particularly in sensitive domains such as health. Structured outputs should not be used to identify, profile, or make decisions about individual users, and downstream users should consider data minimization, access restrictions, and aggregation where appropriate. The attributes discovered by the framework may also reflect biases present in the underlying communities and in the language models used for extraction, and should not be interpreted as representative of broader populations.

Our evaluation focuses on aggregate methodological performance rather than conclusions about individual Reddit users. The running example presented in the paper uses synthetically generated posts to illustrate the extraction and aggregation process. More broadly, we view the proposed framework as a tool for corpus-level research and analysis rather than for individual-level inference or decision making.

% Limitations
% future work

%\bibliography{references.bib}

% If your work needs an appendix, add it before the
% ``\verb|\end{document}|'' command at the conclusion of your source
% document.

% Start the appendix with the ``\verb|appendix|'' command:
% \begin{verbatim}
%   \appendix
% \end{verbatim}
% and note that in the appendix, sections are lettered, not
% numbered. This document has two appendices, demonstrating the section
% and subsection identification method.





\bibliographystyle{ACM-Reference-Format}
\bibliography{bibliography}

@inproceedings{sui2024table,
  title={Table meets llm: Can large language models understand structured table data? a benchmark and empirical study},
  author={Sui, Yuan and Zhou, Mengyu and Zhou, Mingjie and Han, Shi and Zhang, Dongmei},
  booktitle={Proceedings of the 17th ACM International Conference on Web Search and Data Mining},
  pages={645--654},
  publisher = {Association for Computing Machinery},
  address   = {New York, NY, USA},
  year={2024}
}

@article{fennig2025bridging,
  title={Bridging the conversational gap in epilepsy: Using large language models to reveal insights into patient behavior and concerns from online discussions},
  author={Fennig, Uriel and Yom-Tov, Elad and Savitzky, Leehe and Nissan, Johnatan and Altman, Keren and Loebenstein, Roni and Boxer, Marina and Weinberg, Nitai and Gofrit, Shany Guly and Maggio, Nicola},
  journal={Epilepsia},
  volume={66},
  number={3},
  pages={686--699},
  year={2025},
  publisher={Wiley Online Library}
}

@article{yomtov2023large,
  title={A large-scale observational analysis of social media data reveals major public misperception of the attainability of drastic weight loss by dieting},
  author={Yom Tov, Elad and Hochberg, Irit},
  journal={Obesity Facts},
  volume={16},
  number={2},
  pages={141--148},
  year={2023},
  publisher={S. Karger AG}
}

@online{openai_gpt4o_docs,
  title        = {GPT-4o Model Documentation},
  author       = {{OpenAI}},
  url = {https://platform.openai.com/docs/models/gpt-4o},
  year         = {2025},
  note         = {Accessed: 2025-12},
  lastaccessed = {December 2025},
}

@book{hart2001pattern,
  title={Pattern classification},
  author={Hart, Peter E and Stork, David G and Duda, RO},
  url={https://www.worldcat.org/oclc/41347061},
  year={2001},
  publisher={Wiley Hoboken},
address = {Hoboken, NJ},
}

@online{explodingtopics_gpt_parameters2,
  author       = {{Exploding Topics}},
  title        = {Number of Parameters in GPT Models},
  year         = {2025},
  url          = {https://explodingtopics.com/blog/gpt-parameters},
  lastaccessed = {December 2025},
  note         = {Reports public estimates ranging from 200B to 1.8T parameters},
}

@inproceedings{minIE,
  author       = {Kiril Gashteovski and
                  Rainer Gemulla and
                  Luciano Del Corro},
  editor       = {Martha Palmer and
                  Rebecca Hwa and
                  Sebastian Riedel},
  title        = {MinIE: Minimizing Facts in Open Information Extraction},
  booktitle    = {Proceedings of the 2017 Conference on Empirical Methods in Natural
                  Language Processing, {EMNLP} 2017, Copenhagen, Denmark, September
                  9-11, 2017},
  pages        = {2630--2640},
  publisher    = {Association for Computational Linguistics},
  year         = {2017},
  url          = {https://doi.org/10.18653/v1/d17-1278},
  doi          = {10.18653/V1/D17-1278},
  address = {Copenhagen, Denmark},
  bibsource    = {dblp computer science bibliography, https://dblp.org}
}

@article{nadeau2007survey,
  title={A Survey of Named Entity Recognition and Classification},
  author={Nadeau, David and Sekine, Satoshi},
  journal={Lingvisticae Investigationes},
  volume={30},
  number={1},
  pages={3--26},
  year={2007}
}

@inproceedings{Text-to-Table,
    title = "Text-to-Table: A New Way of Information Extraction",
    author = "Wu, Xueqing  and
      Zhang, Jiacheng  and
      Li, Hang",
    editor = "Muresan, Smaranda  and
      Nakov, Preslav  and
      Villavicencio, Aline",
    booktitle = "Proceedings of the 60th Annual Meeting of the Association for Computational Linguistics (Volume 1: Long Papers)",
    month = may,
    year = "2022",
    address = "Dublin, Ireland",
    publisher = "Association for Computational Linguistics",
    url = "https://aclanthology.org/2022.acl-long.180/",
    doi = "10.18653/v1/2022.acl-long.180",
    pages = "2518--2533"
}

@article{sarawagi2008information,
  title={Information Extraction},
  author={Sarawagi, Sunita},
  journal={Foundations and Trends in Databases},
  url = {https://www.emerald.com/ftdbs/article-abstract/1/3/261/1324927/Information-Extraction?redirectedFrom=fulltext},
  volume={1},
  pages={261--377},
  year={2008}
}

@incollection{jiang2012information,
  title={Information Extraction from Text},
  author={Jiang, X. and others},
  booktitle={Handbook of Natural Language Processing},
  url = {https://link.springer.com/chapter/10.1007/978-1-4614-3223-4_2},
  publisher={Springer},
  address = {Berlin, Heidelberg},
  year={2012}

}

@article{li2020survey,
  author       = {Jing Li and
                  Aixin Sun and
                  Jianglei Han and
                  Chenliang Li},
  title        = {A Survey on Deep Learning for Named Entity Recognition},
  journal      = {{IEEE} Trans. Knowl. Data Eng.},
  volume       = {34},
  number       = {1},
  pages        = {50--70},
  year         = {2022},
  url          = {https://doi.org/10.1109/TKDE.2020.2981314},
  doi          = {10.1109/TKDE.2020.2981314},
  bibsource    = {dblp computer science bibliography, https://dblp.org}
}

@inproceedings{lample2016neural,
  author       = {Guillaume Lample and
                  Miguel Ballesteros and
                  Sandeep Subramanian and
                  Kazuya Kawakami and
                  Chris Dyer},
  editor       = {Kevin Knight and
                  Ani Nenkova and
                  Owen Rambow},
  title        = {Neural Architectures for Named Entity Recognition},
  booktitle    = {{NAACL} {HLT} 2016, The 2016 Conference of the North American Chapter
                  of the Association for Computational Linguistics: Human Language Technologies,
                  San Diego California, USA, June 12-17, 2016},
  pages        = {260--270},
  address = {San Diego, California},
  publisher    = {The Association for Computational Linguistics},
  year         = {2016},
  url          = {https://doi.org/10.18653/v1/n16-1030},
  doi          = {10.18653/V1/N16-1030},
  bibsource    = {dblp computer science bibliography, https://dblp.org}
}

@inproceedings{devlin2019bert,
  author       = {Jacob Devlin and
                  Ming{-}Wei Chang and
                  Kenton Lee and
                  Kristina Toutanova},
  editor       = {Jill Burstein and
                  Christy Doran and
                  Thamar Solorio},
  title        = {{BERT:} Pre-training of Deep Bidirectional Transformers for Language
                  Understanding},
  booktitle    = {Proceedings of the 2019 Conference of the North American Chapter of
                  the Association for Computational Linguistics: Human Language Technologies,
                  {NAACL-HLT} 2019, Minneapolis, MN, USA, June 2-7, 2019, Volume 1 (Long
                  and Short Papers)},
  pages        = {4171--4186},
  address = {Minneapolis, Minnesota},
  publisher    = {Association for Computational Linguistics},
  year         = {2019},
  url          = {https://doi.org/10.18653/v1/n19-1423},
  doi          = {10.18653/V1/N19-1423},
  bibsource    = {dblp computer science bibliography, https://dblp.org}
}

@Article{bose2021survey,
AUTHOR = {Bose, Priyankar and Srinivasan, Sriram and Sleeman, William C. and Palta, Jatinder and Kapoor, Rishabh and Ghosh, Preetam},
TITLE = {A Survey on Recent Named Entity Recognition and Relationship Extraction Techniques on Clinical Texts},
JOURNAL = {Applied Sciences},
VOLUME = {11},
YEAR = {2021},
NUMBER = {18},
ARTICLE-NUMBER = {8319},
pages = {8319},
URL = {https://www.mdpi.com/2076-3417/11/18/8319},
ISSN = {2076-3417},
DOI = {10.3390/app11188319}
}

@article{detroja2023survey,
title = {A survey on Relation Extraction},
journal = {Intelligent Systems with Applications},
volume = {19},
pages = {200244},
year = {2023},
issn = {2667-3053},
doi = {https://doi.org/10.1016/j.iswa.2023.200244},
url = {https://www.sciencedirect.com/science/article/pii/S2667305323000698},
author = {Kartik Detroja and C.K. Bhensdadia and Brijesh S. Bhatt}
}

@article{perera2020ner,
  title={Named entity recognition and relation detection for biomedical information extraction},
  author={Perera, Nadeesha and Dehmer, Matthias and Emmert-Streib, Frank},
  journal={Frontiers in cell and developmental biology},
  volume={8},
  pages={673},
  year={2020},
  publisher={Frontiers Media SA}
}

@inproceedings{newsNER,
  author       = {Erik F. Tjong Kim Sang and
                  Fien De Meulder},
  editor       = {Walter Daelemans and
                  Miles Osborne},
  title        = {Introduction to the CoNLL-2003 Shared Task: Language-Independent Named
                  Entity Recognition},
  booktitle    = {Proceedings of the Seventh Conference on Natural Language Learning,
                  CoNLL 2003, Held in cooperation with {HLT-NAACL} 2003, Edmonton, Canada,
                  May 31 - June 1, 2003},
  pages        = {142--147},
  publisher    = {{ACL}},
  year         = {2003},
  address = {Edmonton, Canada},
  url          = {https://aclanthology.org/W03-0419/},
  bibsource    = {dblp computer science bibliography, https://dblp.org}
}

@inproceedings{clinicalNER,
  author       = {Yonghui Wu and
                  Min Jiang and
                  Jun Xu and
                  Degui Zhi and
                  Hua Xu},
  title        = {Clinical Named Entity Recognition Using Deep Learning Models},
  booktitle    = {{AMIA} 2017, American Medical Informatics Association Annual Symposium,
                  Washington, DC, USA, November 4-8, 2017},
  publisher    = {{AMIA}},
  year         = {2017},
  url          = {https://knowledge.amia.org/65881-amiab-1.4254737/t003-1.4258387/f003-1.4258388/2730946-1.4258431/2731659-1.4258428},
  pages   = {1812--1819},
  address = {Washington, DC, USA},
  bibsource    = {dblp computer science bibliography, https://dblp.org}
}

@Article{yang2022survey,
AUTHOR = {Yang, Yang and Wu, Zhilei and Yang, Yuexiang and Lian, Shuangshuang and Guo, Fengjie and Wang, Zhiwei},
TITLE = {A Survey of Information Extraction Based on Deep Learning},
JOURNAL = {Applied Sciences},
VOLUME = {12},
YEAR = {2022},
NUMBER = {19},
ARTICLE-NUMBER = {9691},
pages = {9691},
URL = {https://www.mdpi.com/2076-3417/12/19/9691},
ISSN = {2076-3417},
DOI = {10.3390/app12199691}
}

@INPROCEEDINGS{liu2024structured,
  author={Liu, Xiangfeng and Sun, Jianting and Lei, Aidi and Zhu, Junzhi},
  booktitle={2024 3rd International Conference on Cloud Computing, Big Data Application and Software Engineering (CBASE)}, 
  title={Research and Applications of Large Language Models for Converting Unstructured Data into Structured Data}, 
  year={2024},
  pages={305-308},
  publisher={IEEE},
  address={Hangzhou, China},
  doi={10.1109/CBASE64041.2024.10824634}}

@article{LLMdriventexttotable,
  author       = {Rajmohan C and
                  Sarthak Harne and
                  Arvind Agarwal},
  title        = {{LLM} driven Text-to-Table Generation through Sub-Tasks Guidance and
                  Iterative Refinement},
  journal      = {CoRR},
  volume       = {abs/2508.08653},
  year         = {2025},
  url          = {https://doi.org/10.48550/arXiv.2508.08653},
  doi          = {10.48550/ARXIV.2508.08653},
  eprinttype    = {arXiv},
  eprint       = {2508.08653},
  bibsource    = {dblp computer science bibliography, https://dblp.org}
}

@article{glean,
  author       = {Henry Peng Zou and
                  Siffi Singh and
                  Yi Nian and
                  Jianfeng He and
                  Jason Cai and
                  Saab Mansour and
                  Hang Su},
  title        = {{GLEAN:} Generalized Category Discovery with Diverse and Quality-Enhanced
                  {LLM} Feedback},
  journal      = {CoRR},
  volume       = {abs/2502.18414},
  year         = {2025},
  url          = {https://doi.org/10.48550/arXiv.2502.18414},
  doi          = {10.48550/ARXIV.2502.18414},
  eprinttype    = {arXiv},
  eprint       = {2502.18414},
  bibsource    = {dblp computer science bibliography, https://dblp.org}
}

@inproceedings{ligd,
  author       = {Kangjia Fan and
                  Yilong Zhao and
                  Daifeng Li and
                  Changze Lin and
                  Weijun Zhang and
                  Zhiwen Zhong},
  editor       = {Meeyoung Cha and
                  Chanyoung Park and
                  Noseong Park and
                  Carl Yang and
                  Senjuti Basu Roy and
                  Jessie Li and
                  Jaap Kamps and
                  Kijung Shin and
                  Bryan Hooi and
                  Lifang He},
  title        = {LLM-Enhanced Generalized Category Discovery via Iterative Graph Diffusion},
  booktitle    = {Proceedings of the 34th {ACM} International Conference on Information
                  and Knowledge Management, {CIKM} 2025, Seoul, Republic of Korea, November
                  10-14, 2025},
  pages        = {645--654},
  publisher    = {{ACM}},
  address = {Seoul, Republic of Korea},
  year         = {2025},
  url          = {https://doi.org/10.1145/3746252.3760997},
  doi          = {10.1145/3746252.3760997},
  bibsource    = {dblp computer science bibliography, https://dblp.org}
}

@article{askdocsLLM,
  author = {Sharif, Omar and Basak, Madhusudan and Preum, Sarah M},
  title = {Assessing Large Language Models in Building a Structured Dataset From AskDocs Subreddit Data: Methodological Study},
  journal = {J Med Internet Res},
  year = {2026},
  volume = {28},
  pages = {e74094},
  doi = {10.2196/74094},
  url = {https://www.jmir.org/2026/1/e74094}
}

\end{document}